\documentclass[prl,amssymb,amsmath,twocolumn,aps,showpacs,superscriptaddress,groupedaddress]{revtex4-1}

\usepackage{graphicx}
\usepackage{dcolumn}
\usepackage{bm}
\usepackage{hyperref}

\begin{document}

\title{Onsager coefficients in periodically driven systems}

\author{Karel Proesmans}
 \email{Karel.Proesmans@uhasselt.be}
\author{Christian Van den Broeck}

 \affiliation{Hasselt University, B-3590 Diepenbeek, Belgium.}
 
\date{\today}

\begin{abstract}
We evaluate the Onsager matrix for a system under time-periodic driving by considering all its Fourier components.  By application of the second law, we prove that all the fluxes converge to zero in the limit of zero dissipation. Reversible efficiency can never be reached at finite power.  The implication for an Onsager matrix,  describing reduced fluxes, is that its determinant has to vanish.  In the particular case of only two fluxes, the corresponding Onsager matrix becomes symmetric.  
\end{abstract}

\maketitle
How to reconcile reversibility of microscopic laws with the arrow of time prescribed by the second law of thermodynamics? Ever since the pioneering work of Boltzmann, the question has stirred debate and controversy. Without solving the core issue, Onsager realised that micro-reversibility has implications for macroscopic irreversible laws \cite{onsager1931reciprocal,onsager1931reciprocal2}. In particular, in the regime of linear response around equilibrium, micro-reversibility implies the symmetry of properly defined Onsager coefficients.
Building on preceding case studies \cite{izumida2009onsager,izumida2010onsager,izumida2012efficiency,izumida2015linear,schmiedl2008efficiency,esposito2010quantum}, Brandner et al.~\cite{brandner2015thermodynamics} developed in a recent, remarkable paper the stochastic thermodynamic formalism for periodically driven systems.
In particular, they identify explicitly the Onsager coefficient describing the regime of linear response averaged over one period, and show, amongst other, that they obey a generalized symmetry principle $L_{ij}=\tilde{L}_{ji}$, where the tilde refers to the same system but driven by the time-reversed periodic signal. Under time-symmetric driving, the usual Onsager symmetry  $L_{ij}={L}_{ji}$ is recovered. They next discuss the issue of operating without entropy production, hence reaching reversible efficiency. 
As was pointed out earlier for systems subjected to a magnetic field \cite{benenti2011thermodynamic}, it appears that one could reach such an efficiency while operating at finite power whenever the Onsager matrix is not symmetric. However, a more detailed case study for a magnetic system \cite{brandner2013strong} and one in the Brandner et al.~paper for a system subject to time-asymmetric driving show that ``additional" restrictions appear, which prevent this from occurring, see also \cite{balachandran2013efficiency,brandner2013multi,stark2014classical,sothmann2012magnon,sanchez2015chiral}. It is speculated that these extra conditions lie outside the realm of the second law.
In the present letter we show that this is not the case, provided the full impact of the second law in periodically modulated systems is  assessed.   More precisely, one needs to take into account the symmetries implied by the second law on all the Onsager coefficients that appear,  by considering the decomposition of the periodic perturbation in its Fourier components. 
While, under time-asymmetric periodic driving, asymmetric Onsager matrices may appear in a coarse grained description, the fine structure of the Onsager matrix implies that all fluxes, including the reduced fluxes, converge to zero in the limit of zero dissipation.  The implication for any Onsager matrix, reduced or not, is that its determinant has to vanish in the limit of zero dissipation.  In the particular case of only two fluxes, the corresponding Onsager matrix becomes symmetric. 

We start by reviewing the aforementioned thermodynamic puzzle. Consider a system that is brought out of equilibrium by the application of two thermodynamic forces $F_1$ and $F_2$, for example a temperature gradient and a chemical gradient. As a result, the system is no longer at equilibrium and corresponding fluxes $J_1$ and $J_2$ appear, for example a heat and particle flux. The commonly considered situation is that of a non-equilibrium steady state, so that all these quantities are time independent. For proper choices of the fluxes and forces, this non-equilibrium steady state is characterized by a steady entropy production rate given by:
\begin{eqnarray}\label{ep0}
\dot{S}=F_1J_1+F_2J_2 \geq 0.
\end{eqnarray}
The second law specifies that this quantity cannot be negative. Of special interest is the case in which an entropy decreasing loading process $F_1J_1\leq 0$ (for example a particle flux uphill a chemical gradient), is made possible by an entropy producing driving process  $F_2J_2 \geq 0$. The second law becomes a  statement about the efficiency $\eta$ of this transformation:
\begin{eqnarray}
\eta=-\frac{F_1J_1}{F_2J_2 }\leq 1.\label{EtaDef}
\end{eqnarray}
The maximum efficiency $1$ is obviously reached for a reversible process, i.e.~a  non-dissipative process with zero entropy production. 
For zero forces, the system is at equilibrium and the fluxes vanish. Hence a linear relation is expected between fluxes and forces when the latter are sufficiently small:
\begin{eqnarray}\label{lr}
J_1=L_{11} F_1+L_{12}F_2\nonumber\\
J_2=L_{21} F_1+L_{22}F_2.
\end{eqnarray}
In this region of linear response, the entropy production reduces to the quadratic expression:
\begin{eqnarray}\label{qf}
\dot{S}=L_{11}F^2_1+\left(L_{12}+L_{21}\right)F_1F_2+L_{22}F^2_2.\end{eqnarray}
The second law, requiring the non-negativity of this quadratic form, gives rise to the following conditions:
\begin{eqnarray}\label{csl}
L_{11}\geq0  \;\;,\;L_{22}\geq0\;\;\;
4 L_{11} L_{22}\geq(L_{12}+L_{21})^2.\end{eqnarray}
Zero entropy production is trivially reached for $F_1 = F_2 = 0$. The question of interest is whether this situation, and hence efficiency $1$, can also be realized for non-vanishing values of the forces.
Since the quadratic form Eq.~(\ref{qf}) is non-negative, this can only happen when its zero value is degenerate, i.e., when the discriminant is zero:
\begin{eqnarray}\label{cc}
4 L_{11} L_{22}=(L_{12}+L_{21})^2.\end{eqnarray}
Under this condition, the rate of entropy production can be rewritten as
$\dot{S}=\left(\sqrt{L_{11}}F_1\pm \sqrt{L_{22}}F_2\right)^2$,
and thus vanishes along the line $F_1/F_2=\pm \sqrt{L_{22}/L_{11}}$.  The corresponding efficiency is equal to one. 
Since zero entropy production corresponds to equilibrium, a reasonable guess is that, under this condition, the fluxes should also vanish. 
With the above relation between the forces, one however finds that the fluxes:
\begin{eqnarray}
J_1=\frac{\left(L_{12}-L_{21}\right)}{2}F_2,\;\;\;
J_2=\frac{\left(L_{21}-L_{12}\right)}{2}F_1,\end{eqnarray}
 only vanish for a symmetric Onsager matrix $L_{12}=L_{21}$.  Insisting upon this reasonable line of thought, one concludes that, in the limit of zero entropy production, the $2\times 2$ Onsager matrix has somehow to become symmetric. Another related and revealing observation is to realize from the linear relation Eq.~(\ref{lr}), that vanishing fluxes can only appear for nonzero forces if the determinant of the Onsager matrix is zero, $L_{11} L_{22}=L_{12}L_{21}$. This, combined with a zero discriminant, cf.~Eq.~(\ref{cc}), again implies $L_{12}=L_{21}$.  A zero determinant means that the fluxes are proportional to each other, a property which has been called strong coupling. 
The above  discussion prompts us to ask the following questions.  1) How do we reconcile the above conclusion with the fact that  $2\times 2$ Onsager matrices can be asymmetric?  2) Is the extra condition, rendering the Onsager matrix  symmetric in the zero dissipation limit, extraneous to the second law since it does not follow from Eq.~(\ref{csl})? 3) What about larger Onsager matrices?
 We are able to answer all these questions by investigating in more detail the newly developed thermodynamic theory for systems subject to a time-periodic perturbation  \cite{brandner2015thermodynamics}.  
 The answers, which will be developed in further detail below, are the following. 1) As was already anticipated by another thermodynamic argument in \cite{van2010many}, an Onsager matrix may be asymmetric but has to reduce to a symmetric form in the limit of zero dissipation. A crucial insight is that the reduced Onsager coefficients depend on the type of driving. The coefficients change as one adapts the driving to move closer to the zero dissipation regime. It is in this very limit that the $2\times 2$ Onsager matrix has to be symmetric.  2) From the expression of the entropy production in terms of all the fluxes in response to every possible Fourier mode, we find that zero dissipation implies that all fluxes vanish, and hence also all reduced or  macroscopic fluxes. In particular, any reduced $2\times 2$ Onsager matrix will become symmetric in this limit.  Hence this property is a result of the second law, provided its full impact for  time-periodic perturbations is assessed. 3) The general implication on any Onsager matrix is that its determinant has to go to zero in the zero dissipation limit. For matrices of order higher than two, it does not need to reduce to a symmetric form. 

%

To proceed to these answers, we turn to the stochastic thermodynamics \cite{harris_fluctuation_2007,tome2015stochastic,SpinneyFord,seifert_stochastic_2012,van_den_broeck_ensemble_2014} for a system described by a set of discrete energy levels subject to time-periodic modulation. At this point, we do not need to specify the origin of this modulation. Suffice to say that the modulation could be produced by several sources, which are supposed to be non-dissipative and thus describe the exchange of work between the system and these sources. The system is furthermore in contact with a heat bath at temperature $T$,  which can induce transitions between the different energy levels, entailing the exchange heat with the reservoir. We suppose, without loss of generality, that this temperature is not modulated, but our conclusions apply equally well to, for example, thermal machines.
We denote by $\epsilon_k$ the energy of level $k$.  Its perturbation by a general time-periodic signal  of period  $\mathfrak{T}$ is represented in terms of  its Fourier decomposition:
\begin{equation}\label{ep}
\epsilon_k(t)=\epsilon_{k}+\sum^{\infty}_{n=1}\sum_{\sigma=s,c}\Delta \epsilon_{(k,n,\sigma)} g_{(n,\sigma)}(t),\end{equation}
with
\begin{eqnarray}g_{(n,s)}(t)=\sin\left(\frac{2\pi n}{\mathfrak{T}}t\right)\;\;\;
g_{(n,c)}(t)=\cos\left(\frac{2\pi n}{\mathfrak{T}}t\right)\label{sc}.\end{eqnarray}
$ \Delta\epsilon_{(k,n,\sigma)}$ is the amplitude of the perturbation, applied to energy level $k$, with Fourier mode $n$, and the $\sigma$ index referring to whether it is a sine ($s$) or cosine ($c$) perturbation. 
The rate of entropy production, averaged over one period (and still denoted, by slight abuse of notation, as $\dot{S}$), is given by:
\begin{equation}
\dot{S}=-\frac{1}{\mathfrak{T}}\int^{\mathfrak{T}}_0 dt \frac{\dot{Q}(t)}{T}=-\frac{Q}{\mathfrak{T}T}
=\frac{1}{\mathfrak{T}}\int^{\mathfrak{T}}_0 dt \frac{ \dot{W}(t)}{T}=\frac{W}{\mathfrak{T}T},\label{e1}
\end{equation}
where $\dot{Q}$ and $\dot{W}$ are the rate of heat and work to the system. In writing Eq.~(\ref{e1}), we  assume that the system is in a periodic steady state. Its energy $U$ returns to the same value after each period, which, combined with the first law,  gives $\Delta U=Q+W=0$ for the integral over one period. The state of the system is described by a probability distribution $\bold{p}(t)=\{p_k(t)\}$, with $p_k(t)$ the probability  to be in energy state $k$ at time $t$. This distribution obeys the following master equation:
\begin{equation}\dot{\bold{p}}(t)=\bold{W}(t)\bold{p}(t).\end{equation} 
Work on the system corresponds to energy dispensed upon moving an occupied energy level. In the present stochastic content, the rate of work is thus given by $\dot{W}(t)=\sum_{k}\dot{\epsilon}_k(t) p_k(t)$. Combined with Eqs.~(\ref{ep}) and (\ref{e1}), this leads to the ``familiar" expression for the entropy production as a sum of forces $F_{\alpha}$ times fluxes $J_{\alpha}$ (using the compact notation $\alpha=(k,n,\sigma)$):
\begin{equation}\dot{S}=\sum_{\alpha}F_{\alpha}J_\alpha,\label{DSDef}\end{equation}
\begin{equation}F_{\alpha}=\frac{\Delta \epsilon_{\alpha}}{T},\;\;\; \;\;J_{\alpha}=\frac{1}{\mathfrak{T}}\int^{\mathfrak{T}}_0 dt\, \dot{g}_{(n,\sigma)}(t)p_k(t).\end{equation}

In the regime of linear response, the fluxes are linear functions of the forces:
\begin{eqnarray}\label{lro}J_\alpha&=&\sum_{\beta} L_{\alpha\beta}F_{\beta},\end{eqnarray}
with the Onsager coefficients given by:
\begin{equation}L_{\alpha\beta}=\left.\frac{\partial J_{\alpha}}{\partial F_{\beta}}\right|_{\bold{F}=\bold{0}}.\end{equation}
with $\alpha=(k,n,\sigma)$ and $\beta=(l,m,\rho)$.  Here  $k$ and $l$ refer to energy levels, $n$ and $m$ to  Fourier modes, and $\sigma$ and $\rho$ to the choice of a sine or cosine perturbation.
Via a short calculation, cf.~\cite{brandner2015thermodynamics} and supplemental material, one finds that the Onsager coefficients are the sum of an  adiabatic  and non-adiabatic contribution:
\begin{equation}L_{\alpha\beta}=L^{ad}_{\alpha\beta}+L^{nad}_{\alpha\beta},\end{equation}
\begin{eqnarray}L^{ad}_{(k,m,\sigma),(l,n,\rho)}&=&\frac{p^{eq}_l\left(p^{eq}_k-\delta_{k,l}\right)}{\mathfrak{T}}\int^{\mathfrak{T}}_0 dt\, \dot{g}_{(m,\sigma)}(t)g_{(n,\rho)}(t),\nonumber\\
L^{nad}_{(k,m,\sigma),(l,n,\rho)}&=&\frac{p^{eq}_l}{\mathfrak{T}}\int^{\mathfrak{T}}_0 dt \int^{\infty}_0d\tau\,\dot{g}_{(m,\sigma)}(t)\dot{g}_{(n,\rho)}(t-\tau) \nonumber\\ &&\textbf{1}_k\exp\left(\bold{W}^{(0)}\tau\right)(\bold{1}_l-\bold{p}^{eq}).\end{eqnarray}
$\bold{1}_l$ is the vector  $(0,...,1,...,0)$ with the value $1$ on the $l$ th position, $\bold{W}^{(0)}=\left. \bold{W}(t)\right|_{\bold{F}=\bold{0}}$ is the time-independent unperturbed evolution operator and $\bold{p}^{eq}$ the corresponding equilibrium distribution:
\begin{equation}p^{eq}_k=\frac{e^{-\frac{\epsilon_{k}}{k_B T}}}{\sum_i  e^{-\frac{\epsilon_{i}}{k_BT}}}.\end{equation}
The above integrals can be performed by inserting the expressions given in Eq.~({\ref{sc}). The adiabatic contribution describes the regime of an infinitely slow perturbation. It is anti-symmetric and hence does not contribute to the entropy production. 
 Concentrating further on the non-adiabatic contribution, for which we use the matrix notation $\boldsymbol{L}$, we first observe the ``Curie principle" for time periodic variation: contributions from different frequencies do not mix.
The Onsager matrix $\bold{L}$  can thus be written as a block diagonal matrix or as a direct sum over all frequencies $n$:
\begin{equation}
\bold{L}=\oplus_n \bold{L}_n.
\end{equation}
The contributions per frequency, $\bold{L}_n$, can further be split into two contributions $\bold{L}_n=\bold{L}_n^{(1)}+\bold{L}_n^{(2)}$, where  $\bold{L}_n^{(1)}$  is diagonal in $\sigma$ and $\rho$, while $\bold{L}_n^{(2)}$ is anti-symmetric and hence does not contribute to the entropy production. Explicitly:
\begin{eqnarray}\bold{L}_n^{(1)}&=& \bold{M}^{(1)}_n\otimes \left[ \begin{array}{cc}
1 & 0 \\
0 & 1 \end{array} \right]_{\sigma,\rho},\nonumber\\\bold{L}_n^{(2)}&=& \bold{M}^{(2)}_n\otimes \left[ \begin{array}{cc}
0 & 1 \\
-1 & 0 \end{array} \right]_{\sigma,\rho},\end{eqnarray}
\begin{eqnarray}M^{(1)}_{n;(k,l)}&=&-2\pi^2 n^2\nonumber\\&& \left(\bold{W}^{(0)}\left(4\pi^2n^2\bold{1}+\mathfrak{T}^2{\bold{W}^{(0)}}^2\right)^{-1}\right)_{k,l}p^{eq}_l,\nonumber\\
M^{(2)}_{n;(k,l)}&=&\delta_{k,l}\frac{p^{eq}_k\pi n}{\mathfrak{T}}\nonumber\\&&-\frac{4\pi^3 n^3\left(4\pi^2n^2\bold{1}+\mathfrak{T}^2{\bold{W}^{(0)}}^2\right)^{-1}_{k,l}p^{eq}_l}{\mathfrak{T}}.\label{MFORMULA}
\end{eqnarray}

We are now ready to discuss the implications of zero entropy production. It
 requires thermodynamic forces $\bold{F}$ obeying $\bold{F}\bold{L}\bold{F}=0$.  From the  explicit structure of the Onsager matrix $\bold{L}$, one finds:
\begin{multline}\label{f0}
\sum_{\alpha,\beta} F_{\alpha}L_{\alpha\beta}F_{\beta}=0 \iff F_{(k,n,\sigma)}=F_{(l,n,\sigma)}\;\; \forall (k,l).\end{multline}
This can be  seen by inspection from Eq.~(\ref{MFORMULA}) by decomposing the part of $\textbf{F}$, related to the energy levels, into the left eigenvectors of $\textbf{W}^{(0)}$. Assuming that this matrix is irreducible, the unit vector $\bold{1}=(1,1,...,1)$ is the only left eigenvector with a non-negative (zero) eigenvalue   (see supplemental material for more details). We conclude that zero entropy production is only compatible with a global modulation of the energy levels. Note that as a result  the probability distribution $\textbf{p}$ has to be time-independent. Hence, we expect that all fluxes will  also be zero. Indeed, one immediately verifies that thermodynamic forces of the above type are zero eigenvectors of the Onsager matrix:
\begin{multline}F_{(k,n,\sigma)}=F_{(l,n,\sigma)}\;\;\forall (k,l)\Rightarrow
J_\alpha=\sum_{\beta} L_{\alpha\beta}F_{\beta}=0.
\label{JTot}\end{multline}
This follows again from Eq.~(\ref{MFORMULA}), with the  observation that $\bold{1}$ is also the right eigenvector of both $\bold{M}^{(1)}$ and $\bold{M}^{(2)}$ with  eigenvalue zero. 

We finally return to the thermodynamic puzzle. As an illustrative example, we investigate on the basis of the preceding analysis the transformation between different sources of work $i$. The thermodynamic force $F_\alpha$, describing the intensity of modulation  of a given energy level, is due to the compound effect of all work mechanisms, which we denote by the index $i$:  
\begin{equation}F_\alpha=\sum_iF_{i,\alpha}.
\label{DetOnS}\end{equation} 
Introducing the thermodynamic forces  $F_i$, prescribing the overall intensity of the driving $i$, and the corresponding
  reduced (coarse grained or macroscopic) fluxes:
\begin{equation}\label{ored}J_i=\frac{\sum_\alpha F_{i,\alpha}J_\alpha}{F_i},\end{equation}
one readily sees that the entropy production (\ref{DSDef}) assumes  the  reduced form $\dot{S}=\sum_iF_iJ_i$.
The corresponding reduced Onsager coefficients are obtained by the combination of Eqs. (\ref{lro}) and (\ref{ored}):
\begin{equation}L_{ij}=\sum_{\alpha,\beta}\frac{F_{i,\alpha}F_{j,\beta}}{F_iF_j}L_{\alpha\beta}.\label{MacOn}\end{equation}
We now make the crucial observation  that, in contrast to the detailed Onsager coefficients $L_{\alpha\beta}$, the reduced coefficients $L_{ij}$ do depend on the thermodynamic forces $F_\alpha$. Hence, zero dissipation, which imposes as discussed above conditions on these forces,  will  have an impact on the properties of the reduced Onsager matrix.
Indeed, in the limit of zero dissipation, one concludes from Eq.~({\ref{ored}) that the reduced fluxes $J_i$  go to zero since this is the case for the $J_\alpha$. Referring to our earlier discussion for a $2\times 2$ Onsager matrix, we conclude that  the latter has to become symmetric in the zero dissipation limit. An explicit illustration is given in Fig.~\ref{figex} for a two-level system modulated by two work sources. More generally, the determinant of the Onsager matrix has to vanish (cf.~supplemental material for an example of a reduced $3\times 3$ Onsager matrix, which is clearly asymmetric but has zero determinant). These properties can also be verified directly from the explicit expression Eq.~(\ref{MacOn}).


\begin{figure}\begin{centering}
\includegraphics[width=8cm]{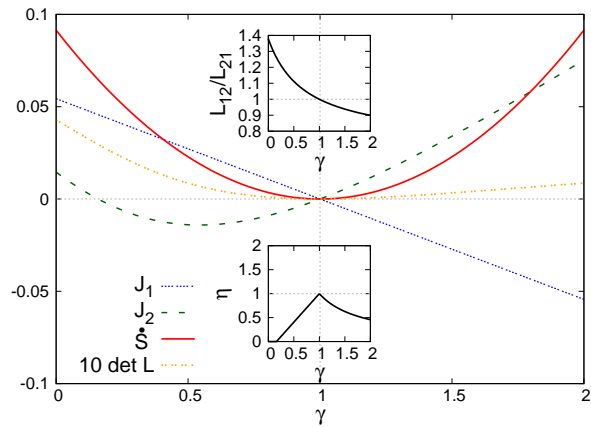}
\caption{A two level system is modulated by two work sources $1$ and $2$, each operating on a different energy level of the system, namely $\epsilon_1(t)=\epsilon_1+\sin(2\pi t)+\cos(2\pi t)$ and $\epsilon_2(t)=\epsilon_2+\sin(2\pi t)+\gamma\cos(2\pi t)$, respectively. The parameters are chosen such that $p^{eq}_1=3/4$, $p^{eq}_2=1/4$, $T=1$, and nonzero eigenvalue of $\textbf{W}^{(0)}$ equal to $-1$.  Main figure: $J_1$ (blue) and  $J_2$ (green)  are the powers of engines $1$ and $2$, respectively, $\dot{S}$ (red) the entropy production, and $10\det L$ (orange) the determinant of the Onsager matrix. $\gamma>1$ and $0.16<\gamma<1$ correspond to different operational regimes with $2$, respectively $1$, functioning as driving and $1$, respectively $2$, as load. The corresponding efficiencies are  $\eta=-F_1J_1/(F_2J_2)$ and $\eta=-F_2J_2/(F_1J_1)$. For $\gamma<0.16$ both fluxes are larger than zero, and the efficiency is given by $\eta=0$. Upper and lower inset: Onsager asymmetry $L_{12}/L_{21}$ and efficiency $\eta$ in function of the parameter $\gamma$. Zero entropy production with reversible efficiency $\eta=1$ is reached for $\gamma=1$, in which case the Onsager matrix becomes symmetric. \label{figex}}
\end{centering}\end{figure}

In conclusion, we showed that zero dissipation implies the vanishing of all fluxes, implying in turn that any Onsager matrix must have a zero determinant.  In particular, a $2\times 2$ Onsager matrix has to become symmetric. It should pose no problem to verify this prediction in a linear response experiment, involving time-periodic perturbation or  magnetic fields, by moving closer to the zero dissipation regime. 
While the analysis given here does not readily apply to magnetic fields,  we surmise that an analogous explanation will hold, probably requiring a more detailed thermodynamic analysis of electromagnetic phenomena.  
Finally, we mentioned that our analysis applies equally well to thermal machines. In fact, the illustrative example from \cite{brandner2015thermodynamics} provides a test case for our predictions: one easily verifies that heat and work flux become proportional to each other in the limit of reversible, i.e. Carnot efficiency, while the Onsager matrix becomes symmetric, cf.~Eqs.~(74) and (72) in this paper \cite{comment}.
\\\\
\begin{acknowledgments}
We thank Thijs Becker and Bart Cleuren for a carefull reading of the manuscript.
\end{acknowledgments}

\bibliographystyle{apsrev4-1}

\begin{thebibliography}{22}
\bibitem{comment}
Due to a transcription error traced back to the appendix, Eq.~D(16), the factors $(1-\bar{\eta})$ and $\bar{\eta}$ have to be interchanged in Eq.~(72). As a result the work and heat protocol become proportional to each other for  $\bar{\eta}=1$
\end{thebibliography}

\end{document}


\title{Onsager coefficients in periodically driven systems}

\author{Karel Proesmans}
 \email{Karel.proesmans@uhasselt.be}
 \affiliation{Hasselt University, B-3590 Diepenbeek, Belgium.}
\author{Christian Van den Broeck}
\affiliation{Hasselt University, B-3590 Diepenbeek, Belgium.}%

\date{\today}
\pacs{05.70.Ln, 05.40.-a}
\maketitle
\section{Adiabatic and non-adiabatic Onsager coefficients: supplemental materials}
The Onsager coefficients are given by:
\begin{eqnarray}L_{(k,m,\sigma),(l,n,\rho)}&=&\left.\frac{\partial J_{(k,m,\sigma)}}{\partial F_{(l,n,\rho)}}\right|_{\boldsymbol{F}=0}\nonumber \\&=&\frac{1}{\mathfrak{T}}\int^{\mathfrak{T}}_0 dt\, \dot{g}_{(m,\sigma)}(t)\left.\frac{\partial p_k(t)}{\partial F_{(l,n,\rho)}}\right|_{\boldsymbol{F}=0}.\label{LApp}\end{eqnarray}
The master equation:
\begin{equation}\dot{\boldsymbol{p}}(t)=\boldsymbol{W}(t)\boldsymbol{p}(t),\end{equation}
can be expanded in the detailed thermodynamic forces:
\begin{equation} \label{mea}\dot{\boldsymbol{p}}_{\alpha}(t)=\boldsymbol{W}^{(0)}\boldsymbol{p}_{\alpha}(t)+\boldsymbol{W}^{(1)}_\alpha (t)\boldsymbol{p}^{eq},\end{equation}
with 
\begin{equation}\boldsymbol{p}_\alpha (t)=\left.\frac{\partial }{\partial F_{\alpha}}\boldsymbol{p}(t)\right|_{\boldsymbol{F}=0},\end{equation}
and
\begin{equation}
\boldsymbol{W}^{(1)}_\alpha(t)=\left.\frac{\partial}{\partial F_\alpha}\boldsymbol{W}(t)\right|_{\boldsymbol{F}=0}.
\end{equation}
The linear Eq.~(\ref{mea}) can be solved explicitly. The contribution of the initial value disappears  when reaching the  time-periodic steady state:
\begin{equation}\boldsymbol{p}_\alpha(t)=\int^{\infty}_0 d\tau\,\exp\left(\boldsymbol{W}^{(0)}\tau\right) \boldsymbol{W}^{(1)}_\alpha(t-\tau) \boldsymbol{p}^{eq}.\end{equation}
It is convenient to introduce the instantaneous steady state of the full operator, $\boldsymbol{p}^{ad}(t)$, which is reached when the system is driven adiabatically slow:
\begin{equation}\label{eqa} \boldsymbol{W}(t)\boldsymbol{p}^{ad}(t)=0,\end{equation}
explicitly:
\begin{equation} \label{adp} p^{ad}_k(t)=\frac{e^{-\frac{\epsilon_k(t)}{k_BT}}}{\sum_i e^{-\frac{\epsilon_i(t)}{k_BT}}}. \end{equation}
By expanding Eq.~(\ref{mea}) in the thermodynamic forces, we arrive at the following relation:
\begin{equation}\boldsymbol{W}^{(0)}\boldsymbol{p}^{ad}_{\alpha}(t)+\boldsymbol{W}^{(1)}_\alpha(t) \boldsymbol{p}^{eq}=0.\end{equation}
This allows to rewrite the quantities $\boldsymbol{p}_\alpha(t)$ in terms of the unperturbed operator and the known adiabatic probabilities, Eq.~(\ref{adp}):
\begin{eqnarray}\boldsymbol{p}_\alpha(t)&=&-\int^{\infty}_0d\tau\,\exp\left(\boldsymbol{W}^{(0)}\tau\right) \boldsymbol{W}^{(0)}\boldsymbol{p}^{ad}_{\alpha}(t-\tau)\nonumber\\&=&\boldsymbol{p}^{ad}_{\alpha}(t)-\int^{\infty}_0d\tau\,\exp\left(\boldsymbol{W}^{(0)}\tau\right) \dot{\boldsymbol{p}}^{ad}_{\alpha}(t-\tau),\end{eqnarray}
where we integrated by parts with respect to $\tau$ to obtain the second line. Returning to equation (\ref{LApp}), we see that:
\begin{eqnarray}L_{(k,m,\sigma),(l,n,\rho)}&=&\frac{1}{\mathfrak{T}}\int^{\mathfrak{T}}_0 dt\, {\dot{g}_{(m,\sigma)}(t)p^{ad}_{\alpha,k}(t)}-\frac{1}{\mathfrak{T}}\int^{\mathfrak{T}}_0 dt\, \int^{\infty}_0 d\tau\, {\boldsymbol{1}_k\dot{g}_{(m,\sigma)}(t)\exp\left(\boldsymbol{W}^{(0)}\tau\right) \dot{\boldsymbol{p}}^{ad}_{\alpha}(t-\tau)}\nonumber\\&=&L^{ad}_{(k,m,\sigma),(l,n,\rho)}+L^{nad}_{(k,m,\sigma),(l,n,\rho)}.\end{eqnarray}
where $\boldsymbol{p}^{ad}_{\alpha}(t)$ is given by:
\begin{equation}\boldsymbol{p}^{ad}_{(l,n,\rho)}(t)=p^{eq}_lg_{(n,\rho)}(t)\left(\boldsymbol{p}^{eq}-\boldsymbol{1}_l\right),\label{padfor}\end{equation}
and $\boldsymbol{1}_l=(0,0,...,1,...,0)$.

\section{Explicit expression of the Onsager coefficients}
The adiabatic part of the Onsager matrix follows from:
\begin{equation}\int^{\mathfrak{T}}_0 dt\, g_{(n,\rho)}(t)\dot{g}_{(m,\sigma)}(t)=(1-\delta_{\sigma,\rho})\delta_{n,m}(-1)^{\delta_{\sigma,c}}\pi n,\end{equation}
namely:
\begin{eqnarray}L^{ad}_{(k,m,\sigma),(l,n,\rho)}&=&\frac{p^{eq}_l(p^{eq}_k-\delta_{k,l})(1-\delta_{\sigma,\rho})\delta_{n,m}(-1)^{\delta_{\sigma,c}}\pi n}{\mathfrak{T}}.\end{eqnarray}
For the non-adiabatic part, we first perform the integral over $t$:
\begin{eqnarray}\int^{\mathfrak{T}}_0 dt\, \dot{g}_{(m,s)}(t)\dot{g}_{(n,s)}(t-\tau)&=&\int^{\mathfrak{T}}_0 dt\, \dot{g}_{(m,c)}(t)\dot{g}_{(n,c)}(t-\tau)\nonumber\\&=&2\delta_{n,m}\frac{\pi^2n^2}{\mathfrak{T}}\cos\left(\frac{2\pi n\tau}{\mathfrak{T}}\right),\end{eqnarray}
\begin{eqnarray}\int^{\mathfrak{T}}_0 dt\, \dot{g}_{(m,s)}(t)\dot{g}_{(n,c)}(t-\tau)&=&-\int^{\mathfrak{T}}_0 dt\, \dot{g}_{(m,c)}(t)\dot{g}_{(n,s)}(t-\tau)\nonumber\\&=&2\delta_{n,m}\frac{\pi^2n^2}{\mathfrak{T}}\sin\left(\frac{2\pi n\tau}{\mathfrak{T}}\right).\end{eqnarray}
To perform the $\tau$-intergral, we make a decomposition in right eigenvectors of  $\boldsymbol{W}^{(0)}$:
\begin{equation}\boldsymbol{p}^{eq}-\boldsymbol{1}_l=\sum_{i} a_i\boldsymbol{\lambda}^r_i,\end{equation}
with $\boldsymbol{\lambda}^r_i$ the right eigenvector corresponding to eigenvalue $\lambda_i$. The corresponding left eigenvector shall be denoted by $\boldsymbol{\lambda}^l_i$. The eigenvectors are chosen to be orthonormal: $\boldsymbol{ \lambda}^l_i\boldsymbol{\lambda}^r_j=\delta_{ij}$.
Note that the left eigenvector, $\boldsymbol{\lambda}^l_0$, corresponding to the supposedly unique zero eigenvalue of $\boldsymbol{W}^{(0)}$ fulfills $\boldsymbol{\lambda}^l_0=\boldsymbol{1}$ for all $k$, due to the orthonormality, and therefore:
\begin{equation}\boldsymbol{\lambda}^l_0\boldsymbol{p}^{eq}-\boldsymbol{\lambda}^l_0\boldsymbol{1}_l=\sum_k p^{eq}_k-1=0,\end{equation}
which leads to $a_0=0$.
Hence there is no contribution form the zero eigenvector. From
\begin{equation}\exp\left(\boldsymbol{W}^{(0)}\tau\right)=\sum_i e^{\lambda_i\tau}\boldsymbol{\lambda}^r_i\boldsymbol{ \lambda}^l_i.\end{equation}
and the observation that all the other eigenvalues $\lambda_i$ of $\boldsymbol{W}^{(0)}$ are strictly negative, the integrals over $\tau$ can be performed:
\begin{equation}\int^{\infty}_0d\tau e^{\lambda_i\tau}\sin\left(\frac{2\pi n\tau}{\mathfrak{T}}\right)=\frac{2\mathfrak{T}\pi n}{4\pi^2 n^2+\lambda_i^2\mathfrak{T}^2},\end{equation}
\begin{equation}\int^{\infty}_0d\tau e^{\lambda_i\tau}\cos\left(\frac{2\pi n\tau}{\mathfrak{T}}\right)=-\frac{\mathfrak{T}^2\lambda_i}{4\pi^2 n^2+\lambda_i^2\mathfrak{T}^2}.\end{equation}
This leads to the following explicit expression for the non-adiabatic entropy Onsager coefficients:
\begin{multline}L^{nad}_{(k,m,\sigma),(l,n,\rho)}=-2\delta_{n,m}\frac{\pi^2n^2p^{eq}_l}{\mathfrak{T}}\sum_{i}a_i\boldsymbol{ 1}_k\boldsymbol{\lambda}^r_i\left((1-\delta_{\sigma,\rho})(-1)^{\delta_{\sigma,c}}\frac{2\pi n}{4\pi^2n^2+\lambda_i^2\mathfrak{T}^2}-\delta_{\sigma,\rho}\frac{\mathfrak{T} \lambda_i}{4\pi^2n^2+\lambda_i^2\mathfrak{T}^2}\right).\end{multline}
Now that all integrals have been executed, we can rewrite the result in matrix notation, by noting that :
\begin{equation}f(\lambda_i)\boldsymbol{\lambda}^r_i=f(\boldsymbol{W}^{(0)})\boldsymbol{\lambda}^r_i.\end{equation}

\section{Zero-entropy production}
To identify the conditions for zero entropy production, we  decompose the thermodynamic force vector into a basis, consisting of vectors of the following form:
\begin{equation}\boldsymbol{F}_{(i,n,\sigma)}=\boldsymbol{F}_n\otimes\boldsymbol{\lambda}^l_i\otimes\boldsymbol{\sigma},\end{equation}
with $\boldsymbol{F}_n=\boldsymbol{1}_n$ the 'natural' basis in the frequency space, $\boldsymbol{\sigma}$ the 'natural' basis in the sine/cosine space, and $\boldsymbol{\lambda}^l_i$ the left eigenbasis of $\boldsymbol{W}^{(0)}$. In this way, a general thermodynamic force vector, $\boldsymbol{F}$ can be written as:
\begin{equation}\boldsymbol{F}=\sum_{i,n,\sigma}b_{(i,n,\sigma)}\boldsymbol{F}_{(i,n,\sigma)}.\end{equation}
We thus find for the rate of entropy production:
\begin{eqnarray}\dot{S}&=&\boldsymbol{F}\boldsymbol{L}\boldsymbol{F}\nonumber\\&=&\sum_{i,n,\sigma}b_{(i,n,\sigma)}^2\boldsymbol{\lambda}^l_i\boldsymbol{M}^{(1)}_n\boldsymbol{\lambda}^l_i\nonumber\\&=&-\sum_{i,n,\sigma,k}b_{(i,n,\sigma)}^2\frac{2\pi^2n^2\lambda_i}{4\pi^2 n^2+\mathfrak{T}^2\lambda^2_i}\left(\boldsymbol{1_k}\boldsymbol{\lambda}^l_i\right)^2 p^{eq}_k\nonumber\\&\geq& 0.\end{eqnarray}
This formula immediately shows that $\dot{S}=0$ if and only if:
\begin{equation} b_{(i,n,\sigma)}=\delta_{i,0}b_{(n,\sigma)},\label{acond}\end{equation}
i.e. the dependency on the energy level of the thermodynamic force is given by $\boldsymbol{\lambda}^l_0=\boldsymbol{1}$, hence the driving of the energy levels is uniform.
Furthermore, it is clear that $\boldsymbol{\lambda}^l_0$ is an eigenvector with eigenvalue zero of both $\boldsymbol{M}^{(1)}_n$ and $\boldsymbol{M}^{(2)}_n$ for all $n$. From this, Eq.~(\ref{acond}) is a sufficient condition to have:
\begin{equation}\boldsymbol{J}=\boldsymbol{L}\boldsymbol{F}=0,\end{equation}
which can also be directly seen from Eqs.~(\ref{LApp}) and (\ref{padfor}). We conclude that zero entropy production implies zero detailed thermodynamic fluxes.

\section{Example: two level system with three thermodynamic forces}
\begin{figure}\begin{centering}
\includegraphics[width=8cm]{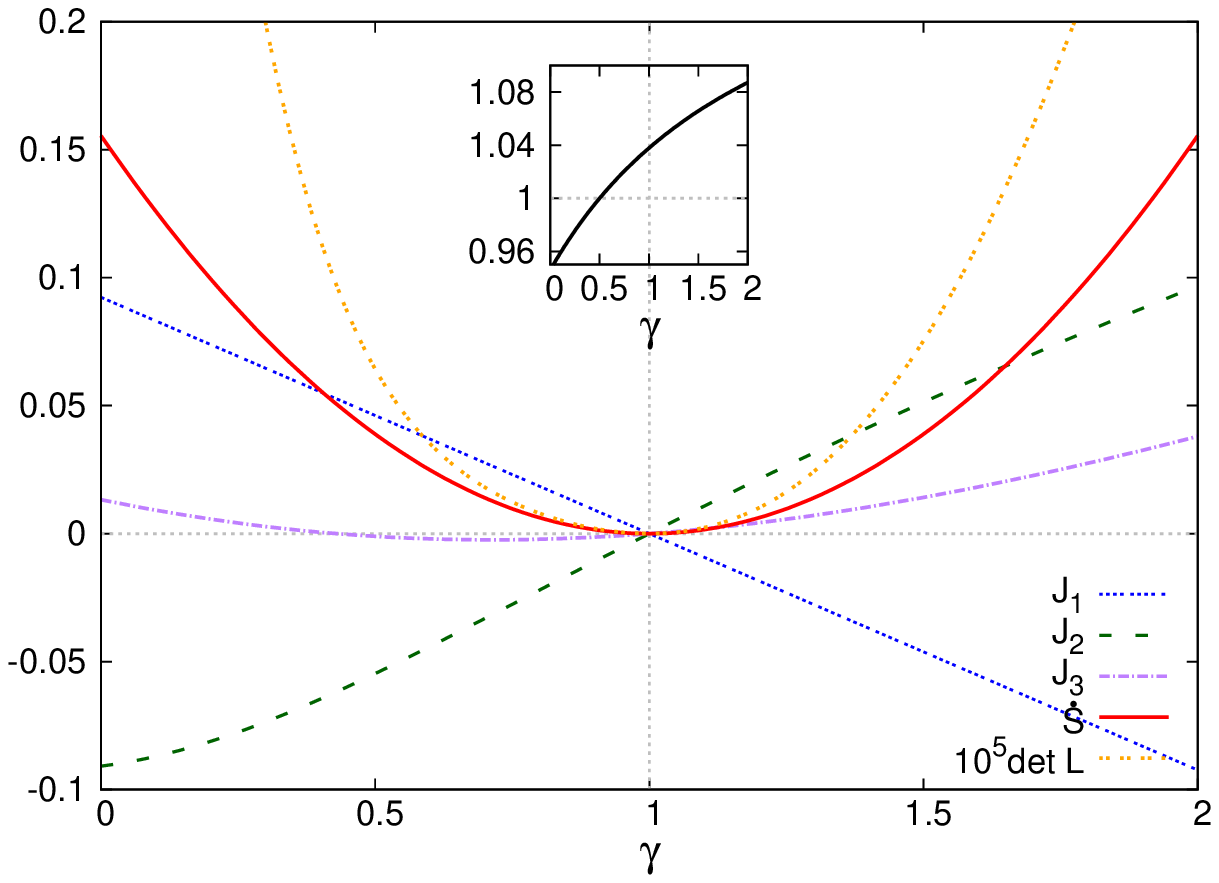}
\caption{Macroscopic fluxes $J_1$ (blue), $J_2$ (green), $J_3$ (purple), the entropy production $\dot{S}$ (red) and $10^5 \det L$ (orange) in function of $\gamma$ with $\gamma_1=\gamma_2=\gamma$. Parameter values: $\mathfrak{T}=1,$ $p^{eq}_1=0.7$,  $p^{eq}_2=0.2,$ $p^{eq}_3=0.1$, $T=1$ $k_1=1$, $k_2=0,$ $k_3=2$. Inset: $L_{12}/L_{21}$ in function of $\gamma$.\label{fig2} }
\end{centering}\end{figure}

In this section we shall consider an extension of the example in the main text. Here we consider a three-level system, where the energy levels are driven as:
\begin{equation}\epsilon_{1}(t)=\epsilon_{1}+F_{(1,s)}\sin\left(\frac{2\pi t}{\mathfrak{T}}\right)+F_{(1,c)}\cos\left(\frac{2\pi t}{\mathfrak{T}}\right),\end{equation}
\begin{equation}\epsilon_{2}(t)=\epsilon_{2}+F_{(2,s)}\sin\left(\frac{2\pi t}{\mathfrak{T}}\right)+F_{(2,c)}\cos\left(\frac{2\pi t}{\mathfrak{T}}\right),\end{equation}
\begin{equation}\epsilon_{3}(t)=\epsilon_{3}+F_{(3,s)}\sin\left(\frac{2\pi t}{\mathfrak{T}}\right)+F_{(3,c)}\cos\left(\frac{3\pi t}{\mathfrak{T}}\right).\end{equation}
In general, $\boldsymbol{W}^{(0)}$ can be written as:
\begin{equation}\boldsymbol{W}^{(0)}=\left[\begin{array}{ccc}-k_1 p^{eq}_2-k_2 p^{eq}_3 & k_1 p^{eq}_1 & k_2 p^{eq}_1 \\ k_1 p^{eq}_2 & -k_1 p^{eq}_1-k_3 p^{eq}_3 & k_3p^{eq}_2\\ k_2 p^{eq}_3 & k_3 p^{eq}_3 & -k_2 p^{eq}_1-k_3 p^{eq}_2\end{array}\right]\end{equation}
We assume that there are three thermodynamic forces, which can be decomposed as:
\begin{equation}F_{1,(1,s)}=2,\;\;\;F_{1,(1,c)}=1,\;\;\;F_{2,(2,s)}=1,\;\;\;F_{2,(2,c)}=\gamma_1,\;\;\;F_{3,(2,s)}=\gamma_2,\;\;\;F_{3,(3,s)}=2,\;\;\;F_{3,(3,c)}=\gamma_2\end{equation}
where $\gamma_1$ and $\gamma_2$ are the variable parameters. The macroscopic forces are then the total amplitudes of the detailed forces:
\begin{equation}F_{1}=\sqrt{5},\;\;\;F_{2}=\sqrt{1+\gamma_1^2},\;\;\;F_{3}=\gamma_2+\sqrt{4+\gamma_2^2},\end{equation}
and the total detailed forces on the energy levels are found by summing up the contributions of the different systems:
\begin{equation}F_{(1,s)}=2,\;\;\;\;F_{(1,c)}=1,\;\;\;\;F_{(2,s)}=1+\gamma_2\;\;\;\;F_{(2,c)}=\gamma_1\;\;\;\;F_{(3,s)}=2\;\;\;\;F_{(2,c)}=\gamma_2.\end{equation}
The results are shown in Fig.~\ref{fig2}. Note that the Onsager matrix is no longer symmetric in the zero dissipation limit, but the determinant still vanishes.
